\documentclass[final,times,sort&compress,onecolumn]{elsarticle}
\usepackage[margin=1in]{geometry}
\usepackage[utf8]{inputenc}
\usepackage{bm}% bold math
\usepackage[dvipsnames]{xcolor}
\usepackage{multicol}
\usepackage{multirow}
\usepackage[normalem]{ulem}
\usepackage{subfig}
\usepackage{graphicx}
\usepackage{amsmath}
\usepackage{slashed}
\allowdisplaybreaks

\renewcommand*{\thefootnote}{\fnsymbol{footnote}}

\begin{document}

\rightline{}

\title{Updated numerical study of transverse single-spin asymmetries in single-inclusive pion production from lepton-nucleon collisions}

\author[label1,label2]{Sophia Fitzgibbons}
\author{Michel Malda$^{a,}$\footnote[2]{Now at L3Harris Technologies, Rochester, New York 14610, USA}}
\author[label1]{Jacob Marsh} 
\author[label1]{Daniel Pitonyak} \ead{pitonyak@lvc.edu}
\author[label1]{Penn Smith} 

\address[label1]{Department of Physics, Lebanon Valley College, Annville, Pennsylvania 17003, USA}
\address[label2]{Department of Physics, SERC, Temple University, Philadelphia, Pennsylvania 19122, USA}

\begin{abstract}
\noindent We revisit the analysis of transverse single-spin asymmetries $A_N$ in lepton-nucleon scattering where only a single pion is detected in the final state, $\ell\,N^\uparrow\to h\, X$.  This observable is the Electron-Ion Collider (EIC) analogue to $A_N$ in proton-proton collisions, $p^\uparrow p\to h\,X$, that has been studied intensely for decades, especially at the Relativistic Heavy Ion Collider (RHIC).  We incorporate new theoretical developments in the collinear twist-3 framework and utilize recent extractions of (Sivers-like and Collins-like) quark-gluon-quark correlators in the numerical computations.  We compare our calculations to HERMES measurements as well as make predictions for Jefferson Lab, COMPASS, and EIC kinematics.  We further explore the role of next-to-leading order (NLO) corrections to the (twist-2) unpolarized cross section (denominator of $A_N$) and consider what can be deduced empirically about the potential numerical significance of the full NLO calculation of $A_N$ in this process.  We consider sources of theoretical uncertainty in our predictions, which present potential opportunities then for future measurements to improve our understanding of $A_N$ and multi-parton correlations in hadrons.

\end{abstract}

\maketitle

\renewcommand*{\thefootnote}{\arabic{footnote}}
\setcounter{footnote}{0}

%%%%%%%%%%%%%%%%%%%%%%%%%%%%%%%%%
%%%%%%%%%%%%%%%%%%%%%%%%%%%%%%%%%
\section{Introduction}
%%%%%%%%%%%%%%%%%%%%%%%%%%%%%%%%%
%%%%%%%%%%%%%%%%%%%%%%%%%%%%%%%%%

The study of transverse single-spin asymmetries (TSSAs) $A_N$ in single-inclusive collisions $pp\to h\,X$ (where one of the incoming protons or the final-state hadron carries a transverse polarization) has played a pivotal role in understanding hadronic structure within perturbative QCD since the late 1970s~\cite{Bunce:1976yb,Klem:1976ui,Kane:1978nd}.  The eventual realization that higher twist (twist-3) multi-parton correlations are central to these processes~\cite{Efremov:1981sh,Efremov:1984ip} paved the way for the development of the collinear twist-3 (CT3) framework often used in calculations of $A_N$ for various reactions~\cite{Qiu:1991pp,Qiu:1991wg,Qiu:1998ia,Kanazawa:2000hz,Eguchi:2006qz,Kouvaris:2006zy,Eguchi:2006mc,Zhou:2008fb,Koike:2009ge,Metz:2012ct,Kanazawa:2013uia,Beppu:2013uda,Kanazawa:2015ajw,Koike:2017fxr,Koike:2019zxc,Koike:2021awj,Koike:2022ddx,Ikarashi:2022yzg,Ikarashi:2022zeo}.  The most recent measurements of $A_N$ in $p^\uparrow p\to h\,X$  are from Fermilab~\cite{Adams:1991rw,Krueger:1998hz} and Brookhaven National Lab~\cite{Allgower:2002qi,Adams:2003fx,Adler:2005in,Lee:2007zzh,Abelev:2008af,Arsene:2008aa,Adamczyk:2012qj,Adamczyk:2012xd,Bland:2013pkt,Adare:2013ekj,Adare:2014qzo,STAR:2020nnl,PHENIX:2023axd}, where especially the Relativistic Heavy Ion Collider (RHIC) has played a pivotal role.  From the theoretical side, focus has been placed on two different types of effects:~initial-state contributions from the ``Sivers-like'' Qiu-Sterman function~\cite{Qiu:1998ia,Kouvaris:2006zy} and final-state contributions from ``Collins-like'' fragmentation correlators~\cite{Metz:2012ct}.\footnote{The full asymmetry also involves terms from so-called soft-fermion poles in the tranversely polarized proton~\cite{Koike:2009ge}, a chiral-odd unpolarized twist-3 correlator~\cite{Kanazawa:2000hz}, and tri-gluon correlators~\cite{Beppu:2013uda}. However, their effects are not numerically significant in the forward region of $p^\uparrow p\to h\,X$~\cite{Kanazawa:2000kp,Kanazawa:2010au,Kanazawa:2011bg,Beppu:2013uda} where large $A_N$ values have been measured.}  Several phenomenological analyses in the CT3 formalism have demonstrated the main cause of $A_N$ to be from the latter~\cite{Kanazawa:2014dca,Gamberg:2017gle,Cammarota:2020qcw,Gamberg:2022kdb}. There are also TSSAs in processes that are sensitive to the intrinsic transverse momentum $k_T$ of partons and require transverse momentum dependent (TMD) factorization~\cite{Collins:1981uk,Collins:1981uw, Collins:1984kg,Meng:1995yn,Idilbi:2004vb,Collins:2004nx,Ji:2004xq,Ji:2004wu,Collins:2011zzd,GarciaEchevarria:2011rb} and the associated parton distribution functions (PDFs) and fragmentation functions (FFs).  These include reactions like semi-inclusive deep-inelastic scattering (SIDIS), electron-positron ($e^+e^-$) annihilation to almost back-to-back hadrons, $pp$ scattering with hadrons detected inside of jets, and Drell-Yan lepton pair or weak gauge boson production -- see Refs.~\cite{Anselmino:2013vqa,Echevarria:2014xaa,Anselmino:2015sxa,Kang:2015msa,Anselmino:2016uie,Kang:2017btw,DAlesio:2017bvu,Echevarria:2020hpy,DAlesio:2020vtw,Bury:2021sue,Bacchetta:2020gko,Cammarota:2020qcw,Gamberg:2022kdb} for some phenomenological analyses.

The future Electron-Ion Collider (EIC) analogue to the $p^\uparrow p\to h\,X$ RHIC measurements is a (na\"{i}vely) simpler reaction theoretically, where the unpolarized proton is replaced by a lepton, i.e., $\ell\,N^\uparrow\to h\, X$, where $N$ is a nucleon.  Experimental data for this process are  already available from HERMES~\cite{Airapetian:2013bim} and Jefferson Lab 6 GeV~\cite{Allada:2013nsw}, although for the latter the hadron transverse momentum is too low to apply perturbative QCD analytical formulas.  Two numerical studies were conducted in 2014, one in the aforementioned CT3 framework~\cite{Gamberg:2014eia} and one in the so-called generalized parton model (GPM)~\cite{Anselmino:2014eza} (see also Refs.~\cite{Anselmino:1999gd,Anselmino:2009pn} for earlier work on $\ell\,N^\uparrow\to h\, X$ in the GPM).  The latter results agreed better with the existing experimental data from HERMES than the former.  However, one has to be careful about a direct comparison between the CT3 and GPM approaches, as they are fundamentally different at a theoretical level, which consequently influences the phenomenology -- see Sec.~IIID of Ref.~\cite{Gamberg:2014eia}.  In 2017 an updated analysis in the GPM was performed~\cite{DAlesio:2017nrd} that included effects in the numerator and denominator of $A_N$ from quasi-real photons through the so-called Weizs\"{a}cker-Williams distribution.  This caused the GPM theoretical calculation to move into even closer alignment with the HERMES data.

Likewise, of particular importance in the CT3 formalism, especially for lower energy fixed-target experiments, is the size of next-to-leading order (NLO) corrections to $A_N$.  For the (twist-2) unpolarized cross section in the denominator of $A_N$ in $\ell\,N^\uparrow\to \pi\, X$, these were calculated in 2015 in Ref.~\cite{Hinderer:2015hra}, where numerically large effects were found.  The NLO computation of the transversely polarized cross section in the numerator, due to its twist-3 nature, is extremely laborious.  Some progress has been made on the Sivers-like contribution~\cite{Voglesang:talkSF17,Marc:privcom}, but a full result that also includes the Collins-like terms is probably many years away. Nevertheless, one may be able to empirically speculate as to the size of NLO corrections in the numerator of $A_N$ in $\ell\,N^\uparrow\to \pi\, X$ by studying the impact of including NLO corrections in the denominator on the ability to describe the HERMES data.  

In addition, there have been two other relevant developments since 2014 with regard to the CT3 approach to understanding the origin of TSSAs.  In 2016, Lorentz invariance relations (LIRs) and equation of motion relations (EOMRs) were derived (or re-derived in some cases) for CT3 PDFs and FFs~\cite{Kanazawa:2015ajw} that, in particular, allow for the Collins-like terms of $A_N$ to be written using only two independent FFs~\cite{Gamberg:2017gle}:~the first $k_T$-moment of the Collins function $H_1^{\perp(1)}(z)$, and the dynamical twist-3 correlator $\tilde{H}(z)$.  Both of those FFs, along with the first $k_T$-moment of the Sivers TMD PDF $f_{1T}^{\perp(1)}(x)$ and transversity collinear PDF $h_1(x)$, were extracted in a 2022 global analysis of TMD and CT3 TSSAs in SIDIS, $e^+e^-$ annihilation, Drell-Yan, and $A_N$ in $p^\uparrow p\to h\,X$~\cite{Gamberg:2022kdb} (see also the work in Ref.~\cite{Cammarota:2020qcw}).  All of these non-perturbative correlators also enter $A_N$ in $\ell\,N^\uparrow\to h\, X$~\cite{Gamberg:2014eia,Kanazawa:2015ajw}.

Given the aforementioned theoretical and phenomenological advances since the original 2014 investigation~\cite{Gamberg:2014eia} of $\ell\,N^\uparrow\to \pi\, X$ in the CT3 framework, as well as the recent community focus on the science of the EIC~\cite{AbdulKhalek:2021gbh}, it is timely to revisit the current status of $A_N$ for this process.  The paper is organized as follows.  In Sec.~\ref{s:theory} we present the relevant analytical formulas for $\ell\,N^\uparrow\to h\, X$.  The numerical implementation of these expressions is discussed in Sec.~\ref{s:pred}, where we make comparisons to the existing HERMES data as well as give predictions for Jefferson Lab (JLab), COMPASS, and the EIC.  We dedicate Sec.~\ref{s:uncert} to a deeper dive into the sources of the theoretical uncertainty in our predictions, and the opportunity then to further understand $A_N$ in $\ell N$ and $pp$ collisions from future measurements.  We conclude and give an outlook in Sec.~\ref{s:concl}.

%%%%%%%%%%%%%%%%%%%%%%%%%%%%%%%%%
%%%%%%%%%%%%%%%%%%%%%%%%%%%%%%%%%
\section{Theoretical Framework} \label{s:theory}
%%%%%%%%%%%%%%%%%%%%%%%%%%%%%%%%%
%%%%%%%%%%%%%%%%%%%%%%%%%%%%%%%%%
The TSSA $A_N$ is generically defined as
\begin{equation}
A_{N} \equiv \frac{\frac{1}{2}\Big\{\!\left[d\sigma_{UT}(\uparrow) - d\sigma_{UT}(\downarrow)\right]\!\Big\}} {d\sigma_{UU}} \equiv \frac{d\sigma_{UT}}{d\sigma_{UU}}\,, \label{e:AN}
\end{equation}
where $d\sigma_{UT}(\vec{S}_T)$ ($d\sigma_{UU}$) is the transverse spin-dependent (unpolarized) cross section, with $\uparrow$ ($\downarrow$) denoting a nucleon with transverse spin $\vec{S}_T$ along the designated positive (negative) transverse axis (e.g., $\pm y$).  Note that we will use $d\sigma_{UT}$ (without the $\vec{S}_T$ argument) to denote the entire numerator of $A_N$.

We consider the reaction $\ell\,N^\uparrow\to \pi\,X$, where the produced final-state pion has a transverse momentum $P_T$.  We define the $+z$-axis to be the direction of $N^\uparrow$'s momentum in the lepton-nucleon center-of-mass (c.m.) frame.  In addition to $P_T$, the asymmetry also depends on the c.m.~energy $\sqrt{S}$ and rapidity $\eta$, which are connected to the Feynman-$x$ variable via $x_F=2P_T\sinh(\eta)/\sqrt{S}$.  The coordinate system is such that at fixed-target experiments like HERMES, JLab, and COMPASS, the final-state pion is produced in the backward region (i.e., negative rapidity/$x_F$).  The two other Mandelstam variables at the hadronic level are 
\begin{equation}
    T= \left(-\sqrt{S}\,\sqrt{P_{T}^2 + x_F^2 S/4}+x_F S/2\right),\quad U=\left(-\sqrt{S}\,\sqrt{P_{T}^2 + x_F^2S/4}-x_F S/2\right).
\end{equation}

The analytical formula for $d\sigma_{UT}^{\ell N^\uparrow \to \pi X}$ at leading order (LO) in the strong coupling $\alpha_s$ and electromagnetic coupling $\alpha_{em}$, i.e., $\mathcal{O}(\alpha_s^0\alpha_{em}^2)$, reads~\cite{Gamberg:2014eia,Kanazawa:2015ajw}
\begin{align}
    d\sigma_{UT}^{\ell N^\uparrow \to \pi X} = \dfrac{-8\alpha_{em}^2P_{T}} {S}\left(\frac{-U}{S^2}\right) &\int_{\frac{U}{T+U}}^{1+\frac{T}{S}}\frac{dv}{v^2(1-v)}\int_{\frac{1-v}{v}\frac{U}{T}}^1\frac{dw}{w^2}\frac{1}{x^2 z^3}\,\delta(1-w)\nonumber\\[0.1cm]
    \times\,&\sum_q e_q^2\left[\frac{M} {\hat{u}}\,D_1^{\pi/q}(z,\mu)\,\mathcal{F}^{q/N}\!(x,\mu,\hat{s},\hat{t},\hat{u}) + \dfrac{M_\pi} {\hat{t}}\,h_1^{q/N}(x,\mu)\,\mathcal{H}^{\pi/q}(z,\mu,\hat{s},\hat{t},\hat{u})\right], \label{e:dsigUT}
\end{align}
where
\begin{align}
\mathcal{F}^{q/N}(x,\mu,\hat{s},\hat{t},\hat{u})&= \left(f_{1T}^{\perp(1)q/N}(x,\mu)-x\frac{df_{1T}^{\perp(1)q/N}(x,\mu)} {dx}\right)\!\left(\frac{\hat{s}(\hat{s}^2+\hat{u}^2)} {2\hat{t}^{\hspace{0.025cm}3}}\right),\label{e:mathcalF}\\[0.3cm]
\mathcal{H}^{\pi/q}(z,\mu,\hat{s},\hat{t},\hat{u})&=\bigg(H_1^{\perp (1) \pi/q}(z,\mu)-z\frac{dH_1^{\perp (1) \pi/q}(z,\mu)} {dz}\bigg)\bigg(\frac{\hat{s}\hat{u}} {\hat{t}^{\hspace{0.025cm}2}}\bigg)+\,\frac{1} {z}\left(-2zH_1^{\perp(1)\pi/q}(z,\mu) +\tilde{H}^{\pi/q}(z,\mu)\right)\!\bigg( \frac{\hat{s} (\hat{u}-\hat{s})} {\hat{t}^{\hspace{0.025cm}2}}\bigg)\,. \label{e:mathcalH}
\end{align}
The non-perturbative inputs in Eqs.~(\ref{e:dsigUT}), (\ref{e:mathcalF}), (\ref{e:mathcalH}) are the unpolarized FF $D_1(z)$, transversity PDF $h_1(x)$~\cite{Ralston:1979ys}, first $k_T$-moment of the Sivers TMD PDF~\cite{Sivers:1989cc,Sivers:1990fh} $f_{1T}^{\perp(1)}(x)$, first $k_T$-moment of the Collins TMD FF~\cite{Collins:1992kk} $H_1^{\perp(1)}(z)$, and the dynamical twist-3 FF $\tilde{H}(z)$~\cite{Kanazawa:2015ajw,Gamberg:2017gle}. The renormalization scale is denoted by $\mu$, and in the numerical computations we use $\mu=P_T$ since $P_T$ sets the hard scale for the process.  The hard factors depend on the partonic Mandelstam variables $\hat{s} = xS, \hat{t}=xT/z, \hat{u}=U/z$, where $x=(1-v)(U/T)/(vw)$ and $z=-T/((1-v)S)$.
The quark fractional charges, in units of the elementary charge $e$, are denoted by $e_q$, with the sum running over active quark and antiquark flavors, and $M$ ($M_\pi$) indicates the nucleon (pion) mass. 

We highlight that Eq.~(\ref{e:dsigUT}) has utilized a LIR and an EOMR among chiral-odd twist-3 FFs~\cite{Kanazawa:2015ajw} to rewrite the fragmentation piece so that it only depends on $H_1^{\perp(1)}(z)$ and $\tilde{H}(z)$.  This form of the analytical result differs from that of the original CT3 analysis of Ref.~\cite{Gamberg:2014eia}, as the aforementioned LIR was derived a couple years after that work, and the current analysis is the first time Eq.~(\ref{e:dsigUT}) has been studied numerically using the most recent simultaneous extractions~\cite{Gamberg:2022kdb} of the relevant PDFs and FFs -- see Sec.~\ref{s:pred}.

The analytical formula for $d\sigma_{unp}^{\ell N \to \pi X}$ up to NLO in the strong coupling $\alpha_s$ and LO in the electromagnetic coupling $\alpha_{em}$, i.e., $\mathcal{O}(\alpha_s\alpha_{em}^2)$, reads~\cite{Hinderer:2015hra}
\begin{align}
d\sigma_{UU}^{\ell N \to \pi X} &= \left(\frac{-U}{S^2}\right)\sum_{i,f}\int_{\frac{U}{T+U}}^{1+\frac{T}{S}}\frac{dv}{v(1-v)}\int_{\frac{1-v}{v}\frac{U}{T}}^1\frac{dw}{w^2}\,\frac{f_1^{i/N}(x,\mu)}{x}\frac{D_1^{\pi/f}(z,\mu)}{z^2}\nonumber\\[0.1cm]
&\hspace{3cm} \times \left[\hat{\sigma}_{\rm LO}^{i\to f}(v)
+\frac{\alpha_s(\mu)}{\pi}\,\hat{\sigma}_{\rm NLO}^{i\to f}(v,w,\mu) + f_{\rm ren}^{\gamma/\ell}\!\left(\!\tfrac{1-v}{1-vw},\mu\right)\,\frac{\alpha_s(\mu)}{\pi}\,\hat{\sigma}_{\rm LO}^{\gamma i\to f}(v,w) \right].\label{e:dsigunp}
\end{align}
The function $f_1(x)$ is the unpolarized PDF, and $f_{\rm ren}^{\gamma/\ell}(y,\mu)$ is the renormalized ``photon-in-lepton'' Weizs\"{a}cker-Williams distribution, which at $\mathcal{O}(\alpha_{em})$ is given by~\cite{Hinderer:2015hra}
\begin{equation}
    f_{\rm ren}^{\gamma/\ell}(y,\mu) = \frac{\alpha_{em}}{2\pi}\,\frac{1+(1-y)^2}{y}\left[\ln\left(\frac{\mu^2}{y^2m_\ell^2}\right)-1\right],
\end{equation}
where $m_\ell$ is the lepton mass.  The sum in Eq.~(\ref{e:dsigunp}) is over partons $i=q\;{\rm or}\; g$ and $f=q\;{\rm or}\; g$.  At LO, the only channel is $q\to q$, and the hard factor is 
\begin{equation}
\hat{\sigma}^{q\to q}_{\rm LO}(v) = \frac{2\alpha_{em}^2}{S}\,e_q^2\left(\frac{\hat{s}^2+\hat{u}^2}{\hat{t}^2}\right)\frac{1}{xv}\,\delta(1-w)\,.
\end{equation}
At NLO, the channels are $q\to q$, $q\to g$, and $g\to q$, with the respective hard factors $\hat{\sigma}^{i\to f}_{\rm NLO}(v,w,\mu)$ given in Eqs.~(26), (27), (28) of Ref.~\cite{Hinderer:2015hra}.  Similarly, the hard factors $\hat{\sigma}_{\rm LO}^{\gamma i\to f}(v,w)$ for the same channels can be found in Eq.~(29) of Ref.~\cite{Hinderer:2015hra}.

%%%%%%%%%%%%%%%%%%%%%%%%%%%%%%%%%
%%%%%%%%%%%%%%%%%%%%%%%%%%%%%%%%%
\section{Numerical Predictions and Comparison with Experimental Data} \label{s:pred}
%%%%%%%%%%%%%%%%%%%%%%%%%%%%%%%%%
%%%%%%%%%%%%%%%%%%%%%%%%%%%%%%%%%
Before presenting our results, we first discuss the numerical implementation of Eqs.~(\ref{e:dsigUT}), (\ref{e:dsigunp}) that allow us to compute $A_N$.  For the unpolarized PDF $f_1(x)$ and FF $D_1(z)$ we utilize CT18 NLO~\cite{Hou:2019qau} and DSS14 NLO~\cite{deFlorian:2014xna}, respectively, using only their central values.  The functions $f_{1T}^{\perp(1)}(x)$, $h_1(x)$, $H_1^{\perp(1)}(z)$, and $\tilde{H}(z)$ are taken from the JAM3D-22 TSSA global analysis of Ref.~\cite{Gamberg:2022kdb}, and we propagate their uncertainties into our computation of $A_N$.  As mentioned previously, JAM3D-22 included TMD and CT3 TSSAs in SIDIS, $e^+e^-$ annihilation, Drell-Yan, and $A_N$ in $p^\uparrow p\to h\,X$.  The predictions for $\ell\, N^\uparrow \to \pi X$ studied here can provide an additional test of that framework.  We utilize the LHAPDF~6.2.3 package~\cite{Buckley:2014ana} to generate input for all the aforementioned non-perturbative functions across the needed momentum fractions and energy scales.  We remind the reader that $f_{1T}^{\perp(1)}(x)$, $H_1^{\perp(1)}(z)$, and $\tilde{H}(z)$ in Eq.~(\ref{e:dsigUT}) can all be written in terms of quark-gluon-quark PDFs/FFs~\cite{Kanazawa:2015ajw}, making $A_N$ in $\ell\, N^\uparrow \to \pi X$ directly sensitive to multi-parton correlations in hadrons.  Therefore, this is an important observable to investigate numerically in order to provide complimentary information to similar CT3 processes like $A_N$ in $p^\uparrow p\to h\,X$~~\cite{Kanazawa:2014dca,Gamberg:2017gle,Cammarota:2020qcw,Gamberg:2022kdb} and $A_{LT}$ in single-inclusive lepton-nucleon and proton-proton collisions~\cite{Kang:2011jw,Kanazawa:2014tda,Bauer:2022mvl}.  

In the following subsections we give our predictions for HERMES, JLab, COMPASS, and EIC kinematics.  For HERMES we compare to their existing experimental measurements, while for the others we give a sample of results that represent the main features of this observable across different c.m.~energies, rapidities, and $P_T$'s.  Plots for any kinematics can be generated upon request.  In all cases we present both results where only the LO unpolarized term (first term in Eq.~(\ref{e:dsigunp})) is included in the denominator of $A_N$ and, for the first time, where the full NLO unpolarized cross section (entire expression in Eq.~(\ref{e:dsigunp})) is used.  Of course a consistent NLO numerical computation also requires NLO terms be included in the numerator, but, as mentioned, such corrections are not available for the transversely polarized (twist-3) cross section.  Nevertheless, by determining the impact of NLO corrections in the unpolarized cross section on the description of the HERMES data, we may be able to empirically deduce the size of the NLO corrections in the numerator of $A_N$.  Given the great complexity of the transversely polarized NLO calculation for this process, any information about their importance is helpful.

As an aside, we mentioned previously an analysis exists in the GPM framework~\cite{DAlesio:2017nrd} where effects from quasi-real photons through the so-called Weizs\"{a}cker-Williams (WW) distribution are included in both the numerator and denominator of $A_N$. We did not pursue such an approach for two reasons.  First, Ref.~\cite{Hinderer:2015hra} showed that the full NLO result for the unpolarized cross section is generally much more sizable than just the WW contribution. Second, since the WW piece is only one part of the full NLO calculation, it is somewhat arbitrary to only keep this term, as there is no guarantee that significant cancellations cannot happen in the $A_N$ numerator with the inclusion of additional terms at NLO that spoil the conclusions one draws by only using the WW term. 

\subsection{Comparison to HERMES Measurements} \label{s:NLO}

\begin{figure}[t!]
\begin{center}
\includegraphics[width=0.71\textwidth]{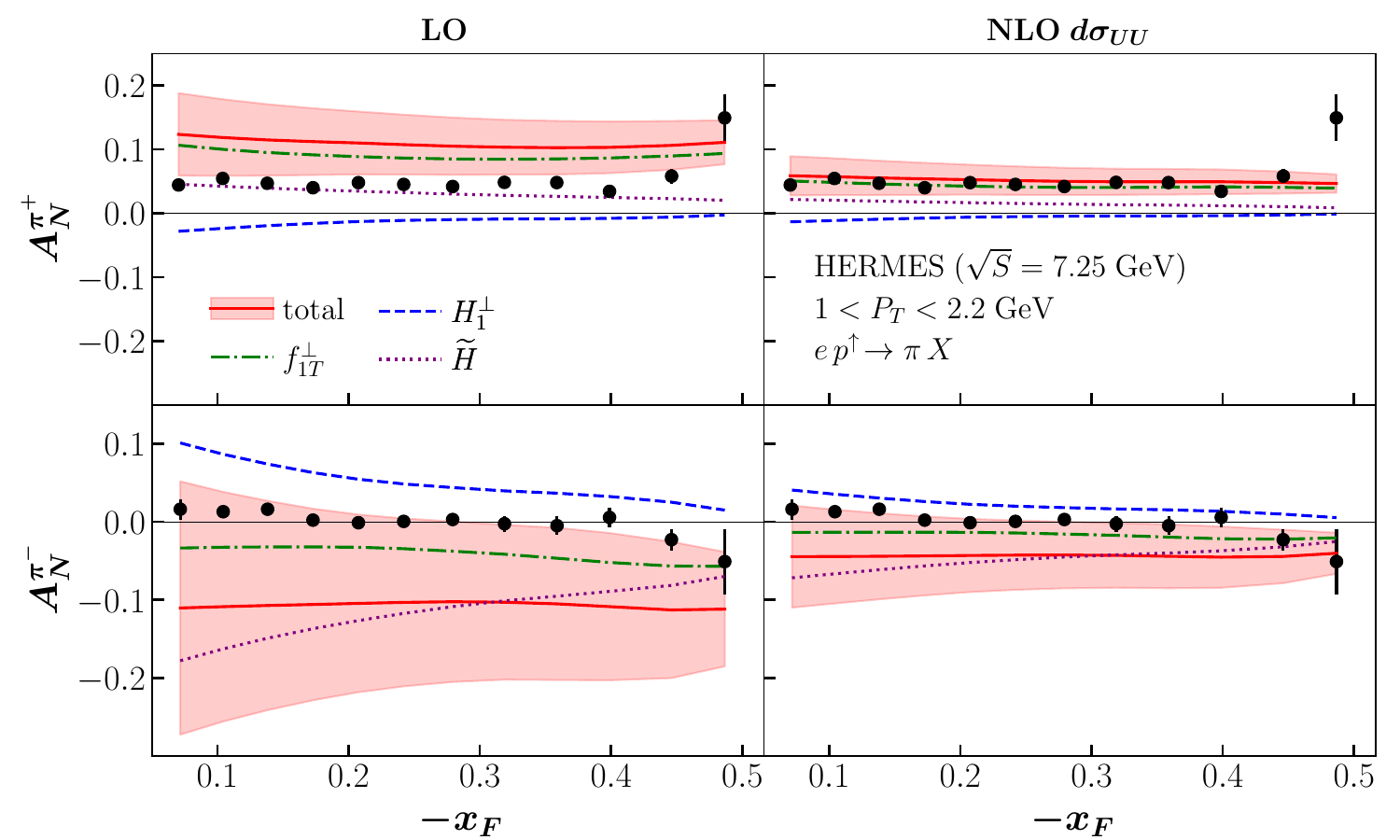}\vspace{-0.4cm}
\end{center}
\caption{Predictions for $A_N$ vs.~$-x_F$ in $ep^\uparrow$ collisions for $\pi^+$ (top row) and $\pi^-$ (bottom row) production compared to the HERMES data $(\sqrt{S}=7.25\,{\rm GeV})$ in Ref.~\cite{Airapetian:2013bim} for the binning $1 < P_T<2.2\,{\rm GeV}$.  The left column gives the LO calculation while the right column includes NLO corrections to the unpolarized cross section (denominator of $A_N$).  The green dot-dashed curve is the contribution to $A_N$ from terms in Eq.~(\ref{e:dsigUT}) involving $f_{1T}^\perp(x)$, the blue dashed curve from terms in Eq.~(\ref{e:dsigUT}) involving $H_1^\perp(z)$, and the purple dotted curve from terms in Eq.~(\ref{e:dsigUT}) involving $\tilde{H}(z)$.  The total result, along with its 1-$\sigma$ uncertainty, is given by the red solid curve and band.} 
\label{f:AN_HERMES_xF}
\end{figure}

In Fig.~\ref{f:AN_HERMES_xF} we compare our  calculation of $A_N$ to HERMES data $(\sqrt{S}=7.25\,{\rm GeV})$ for the $1<P_T<2.2\,{\rm GeV}$ binning as a function of $-x_F$ (recall that fixed-target experiments like HERMES are in the $x_F<0$ region) for both the LO case and also the scenario where we include NLO corrections to the unpolarized cross section (denominator of $A_N$).  We notice that the central curve of the LO result generally is larger in magnitude than the measurements for both $\pi^+$ and $\pi^-$ production, although the outer edge of the error bands overlap with most of the data points.  
%%%%%%%%%%%%%%%%%%%%%%%%%%%%%%%%%%%%%%%%%
%%%%%%%%%%%%%%%%%%%%%%%%%%%%%%%%%%%%%%%%
A similar trend of overshooting the data was observed in Ref.~\cite{Gamberg:2014eia}, except in the present analysis the calculated $\pi^+$ and $\pi^-$ asymmetries are smaller in magnitude than Ref.~\cite{Gamberg:2014eia} (especially for the latter) and have a larger uncertainty. The reason for the larger error bands here is that Ref.~\cite{Gamberg:2014eia} did not propagate uncertainties from the intrinsic and dynamical twist-3 FFs ($H(z)$ and $\hat{H}_{FU}(z,z')$ in that paper), whereas we account for the uncertainty in $\tilde{H}(z)$ (recall that we used a LIR and an EOMR to rewrite $H(z)$ and $\hat{H}_{FU}(z,z')$ in terms of $H_1^{\perp(1)}(z)$ and $\tilde{H}(z)$).  As we explore in more detail in Sec.~\ref{s:uncert}, $\tilde{H}(z)$ is not a very well-constrained function (see Ref.~\cite{Gamberg:2022kdb}) and the term in $A_N$ involving it has large uncertainties.

We consider now the separate contributions (at LO) from the terms in $A_N$ involving $f_{1T}^\perp$, $H_1^\perp$, and $\tilde{H}$. The Sivers-like ($f_{1T}^\perp$) piece is larger for $\pi^+$ than $\pi^-$ and dominates over the Collins-like pieces ($H_1^\perp$ and $\tilde{H}$) in the case of the former.  The $H_1^\perp$ and $\tilde{H}$ terms have opposite signs, with the latter giving the main source of the $\pi^-$ asymmetry, aided by a partial cancellation between the $f_{1T}^\perp$ and $H_1^\perp$ terms.  An interesting observation is that if the $\tilde{H}$ term was negligible for $\pi^-$ production (which corresponds to the unfavored $\tilde{H}(z)$ FF being very small due to $u$-quark dominance from the $e_q^2$ factor in the asymmetry (\ref{e:dsigUT}) and also the fact that the transversity PDF for the $u$ quark is much larger in magnitude than the $d$ quark), then the LO theoretical calculation would be very close to the data.  This highlights the importance of better constraining $\tilde{H}(z)$ in future measurements/analyses.
%%%%%%%%%%%%%%%%%%%%%%%%%%%%%%%%%%%%%%%%%
\begin{figure}[t!]
\begin{center}
\includegraphics[width=0.72\textwidth]{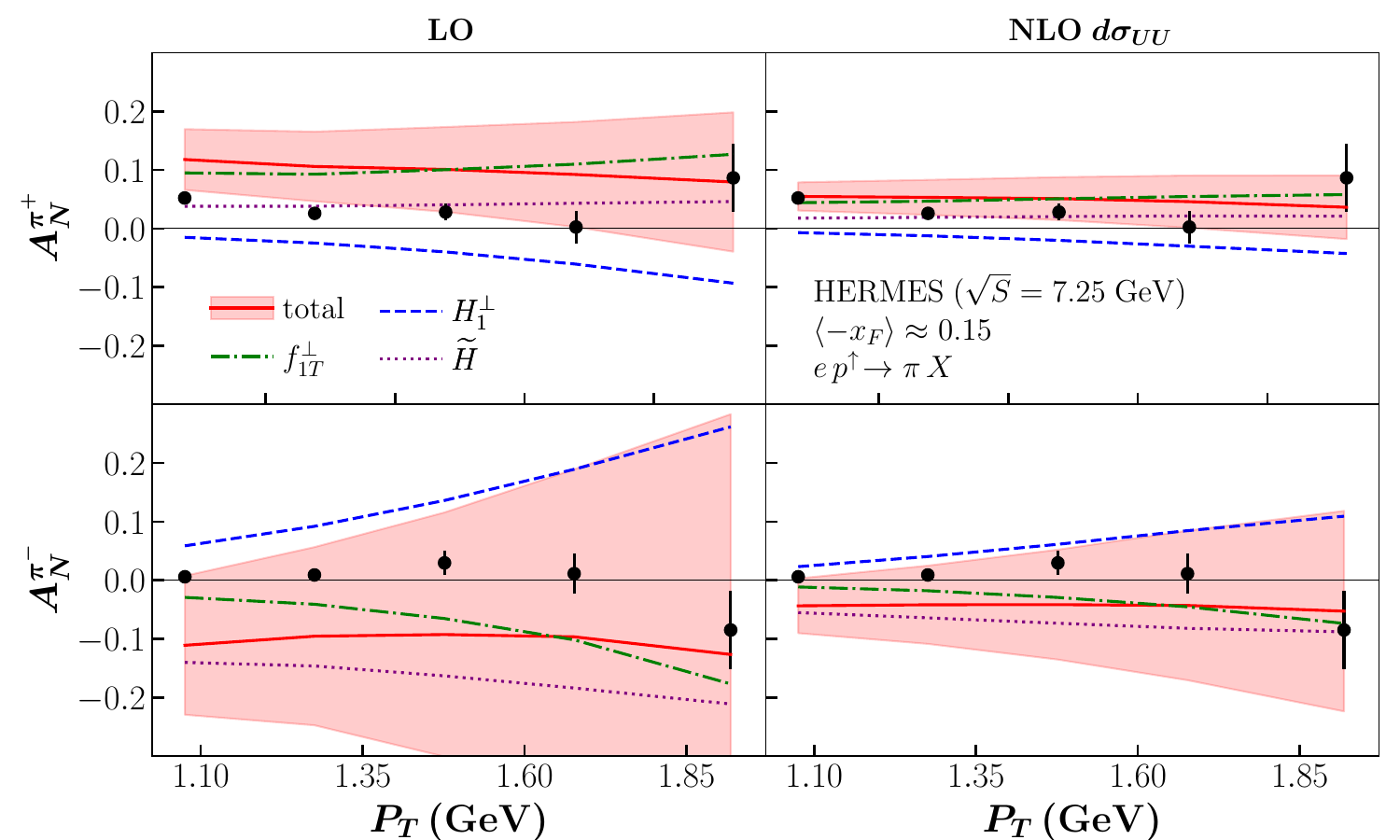} \vspace{-0.4cm}
\end{center}
\caption{Predictions for $A_N$ vs.~$P_T$ in $ep^\uparrow$ collisions for $\pi^+$ (top row) and $\pi^-$ (bottom row) production compared to the HERMES data $(\sqrt{S}=7.25\,{\rm GeV})$ in Ref.~\cite{Airapetian:2013bim} for the binning $0.1 < -x_F<0.2$ ($\langle -x_F \rangle \approx 0.15$).  The left column gives the LO calculation while the right column includes NLO corrections to the unpolarized cross section (denominator of $A_N$).  The green dot-dashed curve is the contribution to $A_N$ from terms in Eq.~(\ref{e:dsigUT}) involving $f_{1T}^\perp(x)$, the blue dashed curve from terms in Eq.~(\ref{e:dsigUT}) involving $H_1^\perp(z)$, and the purple dotted curve from terms in Eq.~(\ref{e:dsigUT}) involving $\tilde{H}(z)$.  The total result, along with its 1-$\sigma$ uncertainty, is given by the red solid curve and band.} 
\label{f:AN_HERMES_PT}
\end{figure}
%%%%%%%%%%%%%%%%%%%%%%%%%%%%%%%%%%%%%%%%

The results where NLO unpolarized corrections are included bring $A_N$ in much better agreement with the data.  This perhaps is not unexpected since we have increased the denominator of $A_N$ without changing the numerator.  The fact, though, that quantitatively the NLO corrections in the denominator alter $A_N$ in this manner did not have to be the case from the outset -- they could have only marginally changed the asymmetry or caused it to become negligible.  We of course cannot say anything definitive about how a full NLO calculation of $A_N$ would compare to the measurements since the NLO corrections in the numerator have not been calculated yet.  Nevertheless, the study performed here may suggest that  NLO corrections to the $A_N$ numerator are small (or need to be small) to account for the HERMES measurements.  
However, typically NLO corrections have similar impact on unpolarized and polarized cross sections so that there is a partial cancellation in the ratio that does not change the corresponding asymmetry significantly.  One example in the context of the process under consideration here is the work in Ref.~\cite{Hinderer:2017ntk}.  The authors analyzed NLO corrections to the (twist-2) longitudinal double-spin asymmetry $A_{LL}$ in $\vec{\ell}\,\vec{N}\to hX$.  Overall, they found the theoretical computation of  $A_{LL}$ at NLO decreased the asymmetry but still was above the available data from SLAC E155.  Of note as well were the predictions for HERMES kinematics did not show a drastic change between LO and NLO.  

This raises the question as to whether it is reasonable to think NLO corrections for $A_N$ in $\ell\, N^\uparrow \to h X$ will behave differently, and if there remain discrepancies with the HERMES data (even with an NLO computation), what this implies about the applicability of the perturbative QCD framework utilized here.  In connection to the former, given the twist-3 nature of $A_N$, and the fact that there are three different types of terms that enter ($f_{1T}^\perp, H_1^\perp, \tilde{H}$), it would not be surprising that there is less cancellation between numerator and denominator at NLO for $A_N$ compared to $A_{LL}$.  That is, the NLO pieces to each individual term ($f_{1T}^\perp, H_1^\perp, \tilde{H}$) may not be small, but perhaps a cancellation occurs between them when added together that makes the overall NLO corrections to the $A_N$ numerator much smaller than the NLO corrections to the $A_N$ denominator.  

If there is still disagreement with the HERMES data, several possible explanations would need to be explored.  Something useful to have is unpolarized cross section data across a wide range of $P_T$ and $\sqrt{S}$ for $\ell N\to hX$ to check if the existing (unpolarized) NLO calculation matches these measurements.  If it does, then one can proceed with more confidence that perturbative QCD can describe $A_N$ even at HERMES kinematics.  The issue then may be the twist-3 PDFs/FFs need to be re-extracted including the $A_N$ data from HERMES (and future experiments) since these functions are not as well constrained as twist-2 PDFs/FFs (see also Sec.~4).  If the unpolarized cross section cannot be described by an NLO calculation, then higher orders may have to be included, or one may need to consider if there is a purely non-perturbative mechanism at play, especially at lower $P_T$ and $\sqrt{S}$.  In either case, more measurements of $\ell N\to hX$ ultimately must be performed before any definitive conclusions can be reached.

In Fig.~\ref{f:AN_HERMES_PT} we compare our calculation of $A_N$ to $P_T$-dependent HERMES data for the $0.1<-x_F<0.2$ binning ($\langle -x_F\rangle \approx 0.15$).\footnote{HERMES also has binnings for $0<-x_F<0.1$, $0.2<-x_F<0.3$, and $0.3<-x_F<0.55$.  The data and theoretical calculation display a similar trend to the $0.1<-x_F<0.2$ binning.}  We notice that the $H_1^\perp$ and $\tilde{H}$ terms largely cancel, leaving the Sivers-like term to mostly dictate the behavior of $A_N$ as a function of $P_T$.  Similar to the $x_F$-dependent result, the LO computation has a greater magnitude than the data, except at the largest $P_T$ value.  Including NLO corrections in the unpolarized cross section bring the theoretical curves in better agreement with the measurements.  Nevertheless, the error bands are such that even the LO calculation mostly overlaps with the measurements.

\subsection{Predictions for JLab and COMPASS}

\begin{figure}[h!]
\begin{center}
\includegraphics[width=0.73\textwidth]{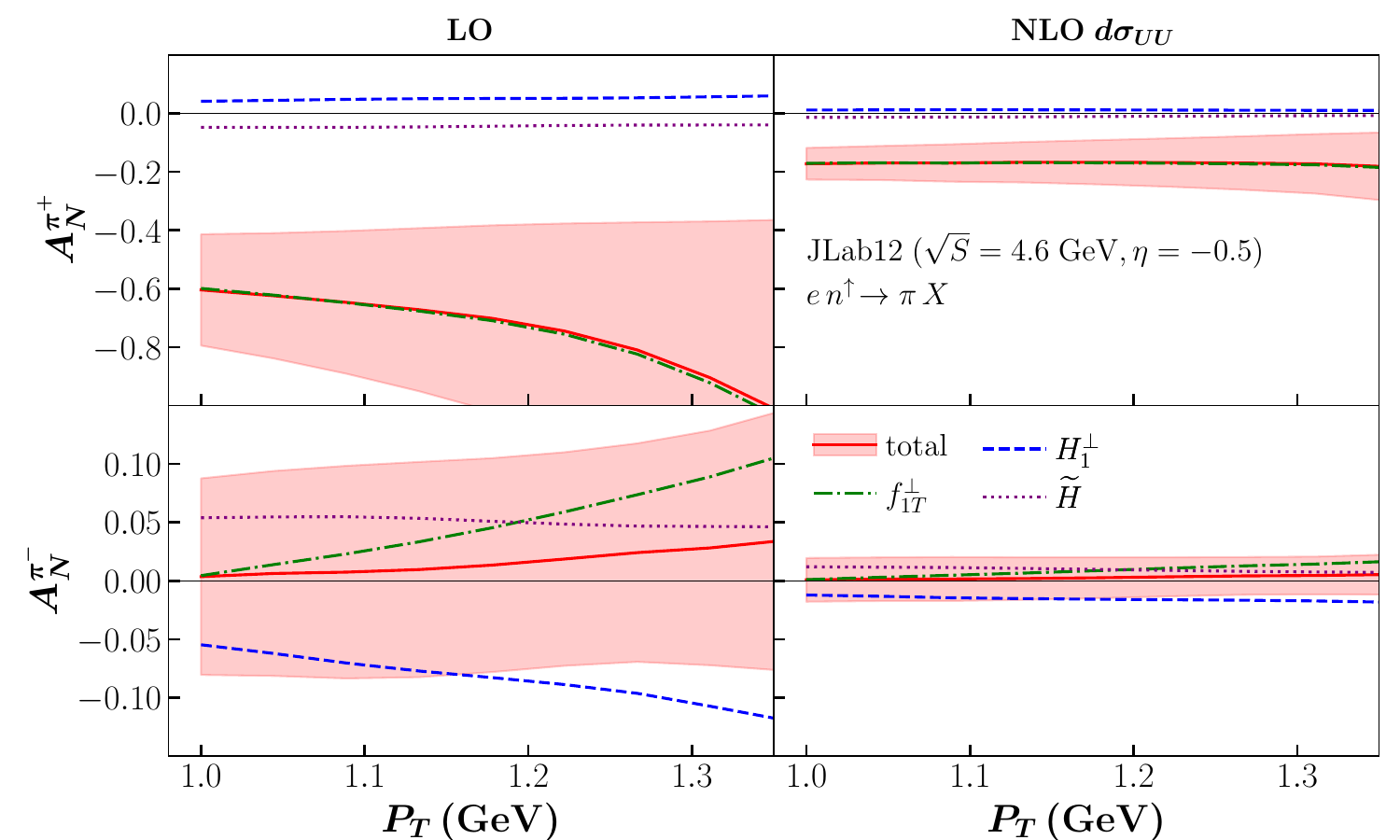} \vspace{-0.4cm}
\end{center}
\caption{Predictions for $A_N$ vs.~$P_T$ in $en^\uparrow$ collisions for $\pi^+$ (top row) and $\pi^-$ (bottom row) production for JLab 12 GeV kinematics $(\sqrt{S}=4.6\,{\rm GeV}, \eta = -0.5)$.  The left column gives the LO calculation while the right column includes NLO corrections to the unpolarized cross section (denominator of $A_N$).  The green dot-dashed curve is the contribution to $A_N$ from terms in Eq.~(\ref{e:dsigUT}) involving $f_{1T}^\perp(x)$, the blue dashed curve from terms in Eq.~(\ref{e:dsigUT}) involving $H_1^\perp(z)$, and the purple dotted curve from terms in Eq.~(\ref{e:dsigUT}) involving $\tilde{H}(z)$.  The total result, along with its 1-$\sigma$ uncertainty, is given by the red solid curve and band.} 
\label{f:AN_JLab12}
\end{figure}

In Fig.~\ref{f:AN_JLab12} we give predictions for $A_N$ in $en^\uparrow\to \pi\,X$ for JLab 12 GeV kinematics ($\sqrt{S}=4.6\,{\rm GeV}, \eta = -0.5$) as a function of $P_T$. 
We find that the LO $\pi^+$ asymmetry can be very large ($70-80\%$), whereas the $\pi^-$ asymmetry shows a much milder behavior ($-10\%$ to $+10\%$).  In both final states there is a cancellation between the $H_1^\perp$ and $\tilde{H}$ terms, leaving $A_N$ to be mainly driven by the Sivers-like term.  When NLO corrections are included in the unpolarized cross section, the $\pi^+$ asymmetry is reduced to $\sim 20\%$ while the $\pi^-$ case becomes negligible.
%%%%%%%%%%%%%%%%%%%%%%%%%%%%%%%%%%%%%%%%%
%%%%%%%%%%%%%%%%%%%%%%%%%%%%%%%%%%%%%%%%
%%%%%%%%%%%%%%%%%%%%%%%%%%%%%%%%%%%%%%%%%
\begin{figure}[t!]
\begin{center}
\includegraphics[width=0.76\textwidth]{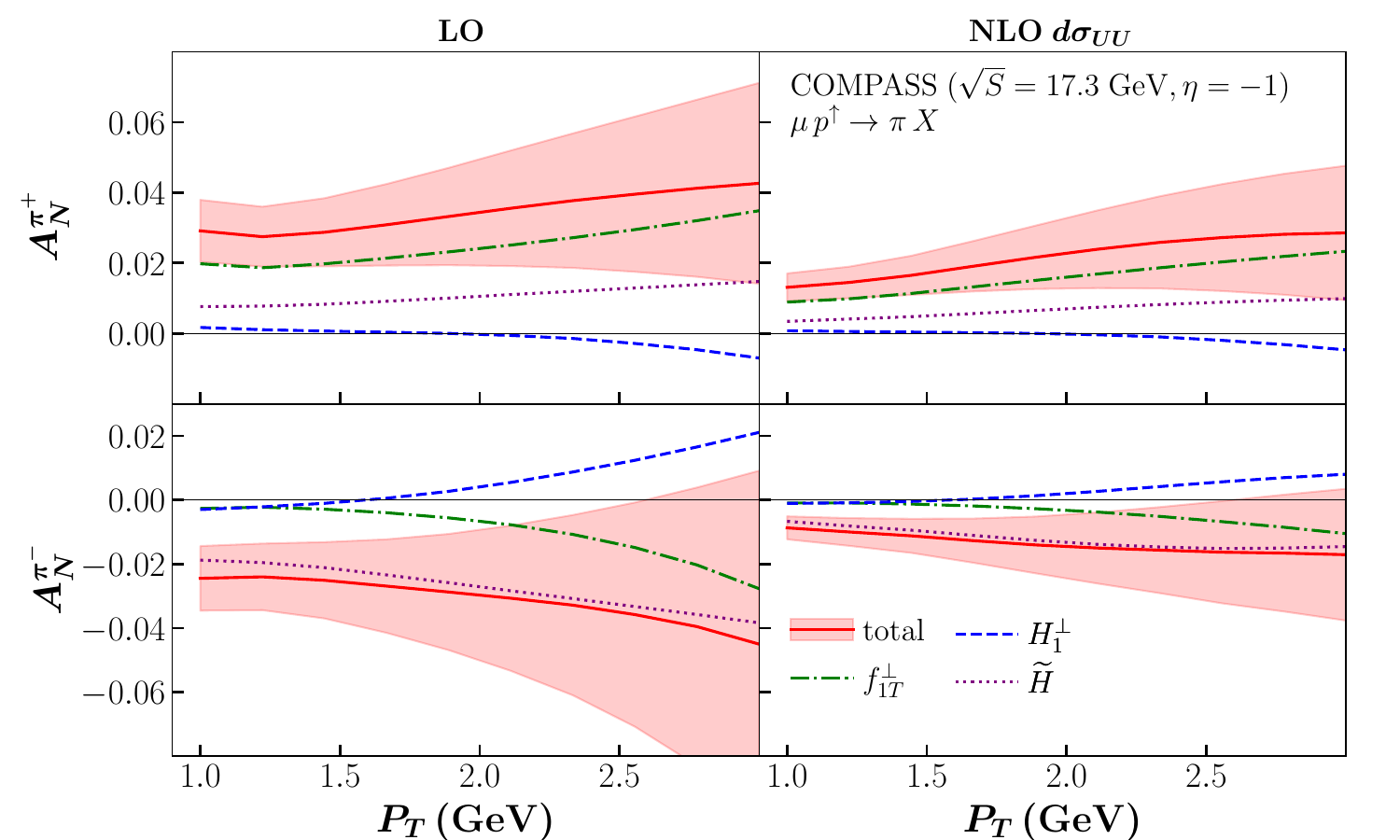} \vspace{-0.4cm}
\end{center}
\caption{Predictions for $A_N$ vs.~$P_T$ in $\mu p^\uparrow$ collisions for $\pi^+$ (top row) and $\pi^-$ (bottom row) production for COMPASS kinematics $(\sqrt{S}=17.3\,{\rm GeV}, \eta = -1)$.  The left column gives the LO calculation while the right column includes NLO corrections to the unpolarized cross section (denominator of $A_N$).  The green dot-dashed curve is the contribution to $A_N$ from terms in Eq.~(\ref{e:dsigUT}) involving $f_{1T}^\perp(x)$, the blue dashed curve from terms in Eq.~(\ref{e:dsigUT}) involving $H_1^\perp(z)$, and the purple dotted curve from terms in Eq.~(\ref{e:dsigUT}) involving $\tilde{H}(z)$.  The total result, along with its 1-$\sigma$ uncertainty, is given by the red solid curve and band.} 
\label{f:AN_COMPASS}
\end{figure}
%%%%%%%%%%%%%%%%%%%%%%%%%%%%%%%%%%%%%%%%

In Fig.~\ref{f:AN_COMPASS} we give predictions for $A_N$ in $\mu p^\uparrow\to \pi\, X$ for COMPASS kinematics ($\sqrt{S} = 17.3\,{\rm GeV}, \eta=-1$) as a function of $P_T$.  We find the $\pi^+$ and $\pi^-$ asymmetries are roughly equal in magnitude ($\sim 3\%$ for the LO calculation and $\sim 1-2\%$ when NLO corrections are included in the unpolarized cross section) and opposite in sign.  For the $\pi^+$ asymmetry, a partial cancellation again occurs between the $H_1^\perp$ and $\tilde{H}$ terms so that $A_N$ follows the trend of the $f_{1T}^\perp$ part.  However, an intriguing feature of $A_N$ for $\pi^-$  is that the $f_{1T}^\perp$ and $H_1^\perp$ terms cancel, allowing the $\tilde{H}$ term to be the main source of the asymmetry.  Therefore, a measurement of this observable could give us further insight into the quark-gluon-quark FF $\tilde{H}(z)$.  More broadly, any data from JLab or COMPASS on $A_N$ in $\ell\,N^\uparrow \to \pi\, X$ would be helpful in trying to better understand the mechanism behind the asymmetry and the role of NLO corrections, both of which will allow for more precise calculations for future EIC experiments.

\subsection{Predictions for the EIC}

\begin{figure}[p!]
\begin{center}
\includegraphics[width=0.71\textwidth]{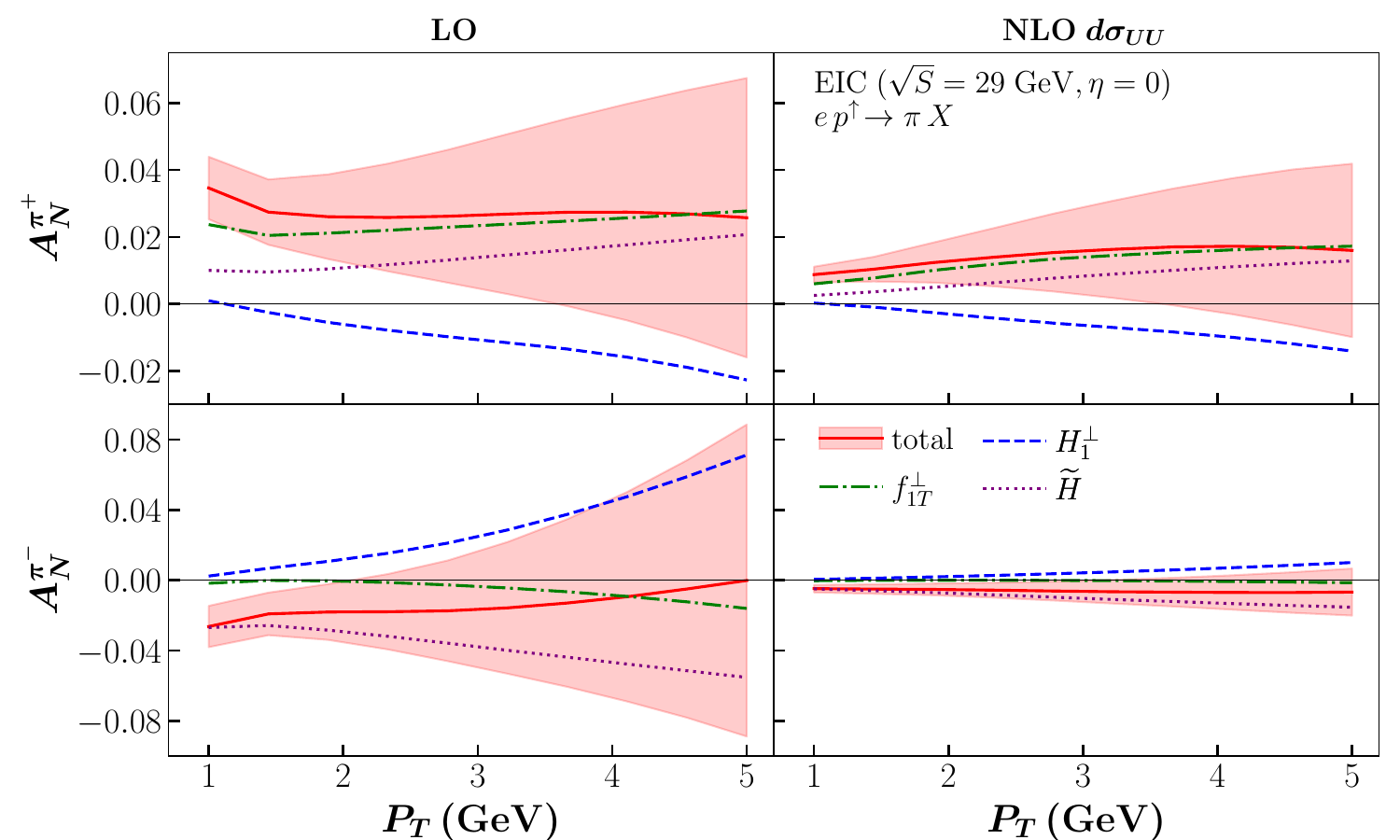}\\[0.05cm]
\includegraphics[width=0.73\textwidth]{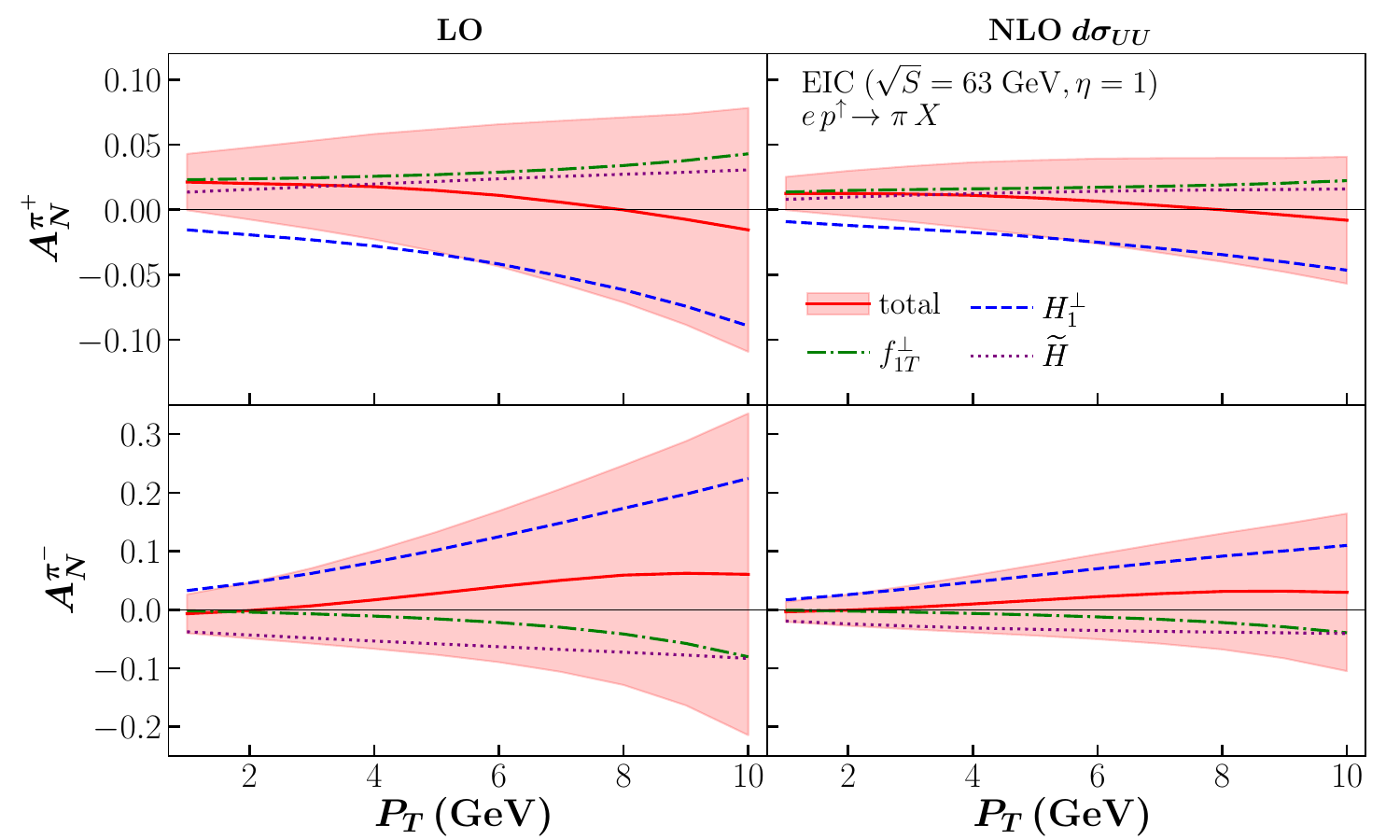}\\[0.05cm]
\includegraphics[width=0.72\textwidth]{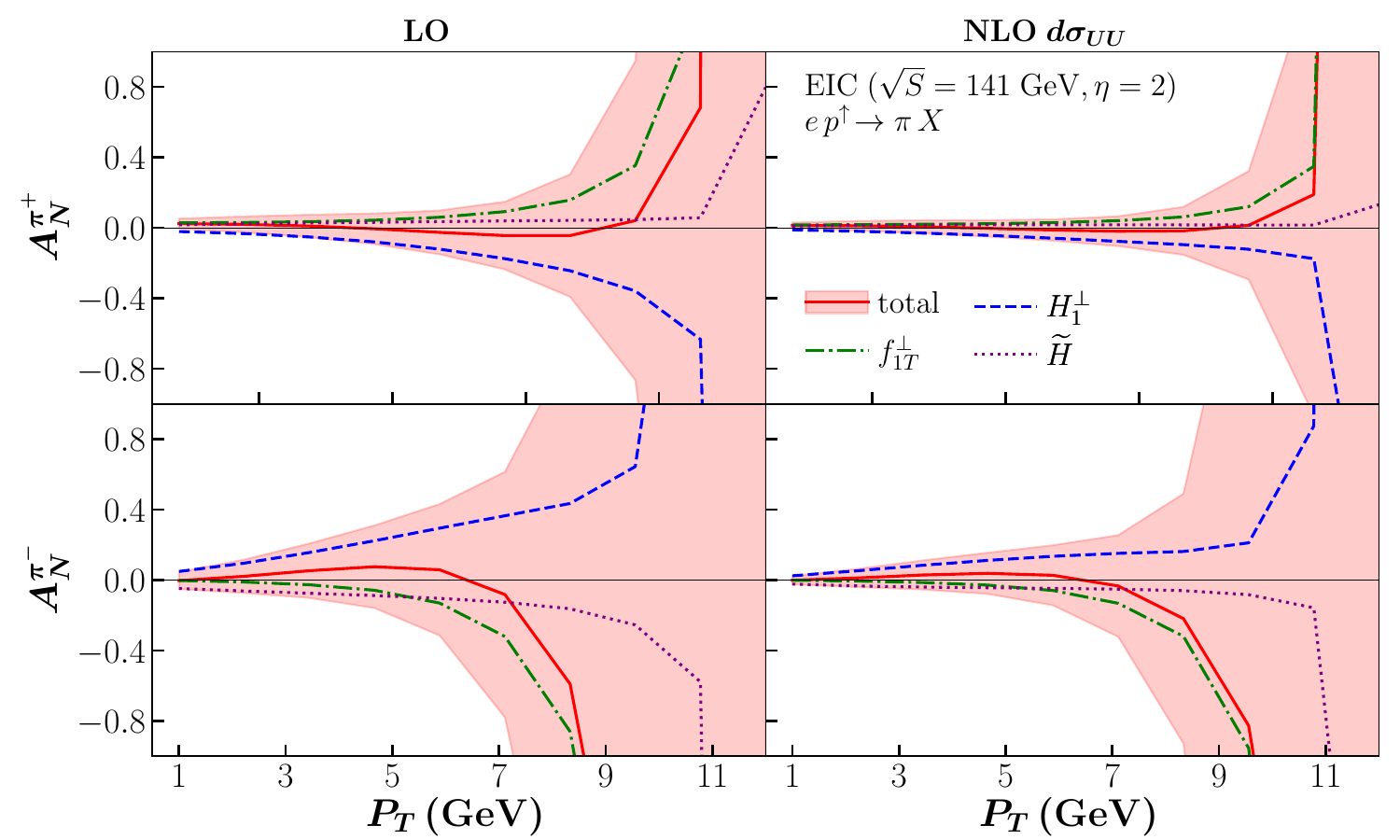} \vspace{-0.4cm}
\end{center}
\caption{Predictions for $A_N$ vs.~$P_T$ in $ep^\uparrow$ collisions for $\pi^+$ (top row of each subfigure) and $\pi^-$ (bottom row of each subfigure) production for various EIC kinematics $(\sqrt{S}=29\,{\rm GeV}, \eta = 0; \sqrt{S}=63\,{\rm GeV}, \eta = 1; \sqrt{S}=141\,{\rm GeV}, \eta = 2)$.  The left column of each subfigure gives the LO calculation while the right column includes NLO corrections to the unpolarized cross section (denominator of $A_N$).  The green dot-dashed curve is the contribution to $A_N$ from terms in Eq.~(\ref{e:dsigUT}) involving $f_{1T}^\perp(x)$, the blue dashed curve from terms in Eq.~(\ref{e:dsigUT}) involving $H_1^\perp(z)$, and the purple dotted curve from terms in Eq.~(\ref{e:dsigUT}) involving $\tilde{H}(z)$.  The total result, along with its 1-$\sigma$ uncertainty, is given by the red solid curve and band.} 
\label{f:AN_EIC}
\end{figure}

In Fig.~\ref{f:AN_EIC} we give predictions for low-, medium-, and high-energy EIC configurations at select rapidities:~$\sqrt{S}=29\,{\rm GeV},\eta =0; \sqrt{S}=63\,{\rm GeV}, \eta=1; \sqrt{S}=141\,{\rm GeV}, \eta=2$, respectively.  These choices give a representative sample of how the asymmetry behaves and where it is most sizable.  Plots of $A_N$ for any kinematics can be generated upon request.  First, for the low-energy scenario we find the LO calculation provides asymmetries $\sim 3\%$ for $\pi^+$ and $\pi^-$ production that have opposite signs and decrease with increasing $P_T$.  The inclusion of NLO corrections in the unpolarized cross section reduce the asymmetries to $\sim 1\%$ in magnitude.  In both the LO and NLO cases the theoretical uncertainties increase at larger $P_T$, which is a general feature of all the EIC scenarios that we will explore in more detail in Sec.~\ref{s:uncert}.  We observe that in the LO result for $\pi^-$, if the $\tilde{H}$ term was negligible, $A_N$ would be positive instead of trending negative.  This again highlights the important role played by the quark-gluon-quark FF $\tilde{H}(z)$ and the need to better constrain this function.
%%%%%%%%%%%%%%%%%%%%%%%%%%%%%%%%%%%%%%%%%
%%%%%%%%%%%%%%%%%%%%%%%%%%%%%%%%%%%%%%%%

Second, for the medium-energy scenario we find at LO that both the $\pi^+$ and $\pi^-$ asymmetries trend positive due to the $f_{1T}^\perp$ and $\tilde{H}$ terms reinforcing each other in the former and the $H_1^\perp$ term being dominant in the latter.  However, the error bands are such that the sign of $A_N$ cannot be definitively determined and a wide range of values are possible ($\sim -10\%$ to $+5\%$ for $\pi^+$ and $\sim -20\%$ to $+30\%$ for $\pi^-$).  The inclusion of NLO corrections to the unpolarized cross section reduces the asymmetries by about a factor of two.

Lastly, for the high-energy scenario we find at LO that $A_N$ for $\pi^+$ is small and consistent with zero up until $P_T \approx 5\,{\rm GeV}$ due to a cancellation between the $f_{1T}^\perp$ and $H_1^\perp$ terms and the $\tilde{H}$ term being negligible. The asymmetry then starts to trend slightly negative until turning very large and positive.  $A_N$ for $\pi^-$ trends positive at smaller $P_T$ (up to $\sim 6\,{\rm GeV}$) due to the $H_1^\perp$ term being larger in magnitude than the other two terms.  The asymmetry then turns large and negative due to the Sivers-like term growing much more rapidly than the Collins-like terms.  The inclusion of NLO corrections in the unpolarized cross section reduces the asymmetries in the small-$P_T$ region but does not meaningfully tame the rise in magnitude of $A_N$ at large $P_T$.  The large increase in the asymmetries at high $P_T$ could be due to the fact that one is moving further into forward kinematics where the PDFs and FFs are being probed at larger momentum fractions that approach 1, and threshold resummation techniques may be required~\cite{Anderle:2012rq,Anderle:2013lka,Hinderer:2014qta,Hinderer:2018nkb,Kaufmann:2019ksh}.  This is also the regime (larger $x$ (and $z$)) where one becomes more sensitive to the fact that the {\it derivatives} of $f_{1T}^{\perp(1)}(x)$ and $H_1^{\perp(1)}(z)$ enter the theoretical formulas (see Eqs.~(\ref{e:mathcalF}), (\ref{e:mathcalH})). Since the functions themselves obey simple power laws in $(1-x)$ (or $(1-z)$), the derivatives may not fall off as quickly in the forward region as the unpolarized PDFs and FFs.

%%%%%%%%%%%%%%%%%%%%%%%%%%%%%%%%%
%%%%%%%%%%%%%%%%%%%%%%%%%%%%%%%%%
\section{Sources of Theoretical Uncertainty and Possible Opportunities} \label{s:uncert}
\begin{figure}[h!]
\begin{center}
\includegraphics[width=0.71\textwidth]{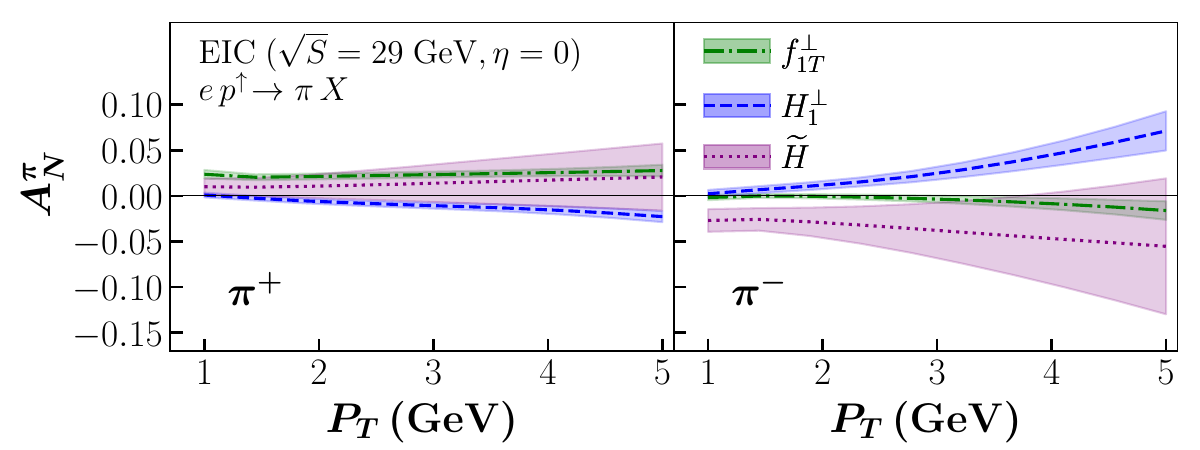}\\[0.1cm]
\includegraphics[width=0.70\textwidth]{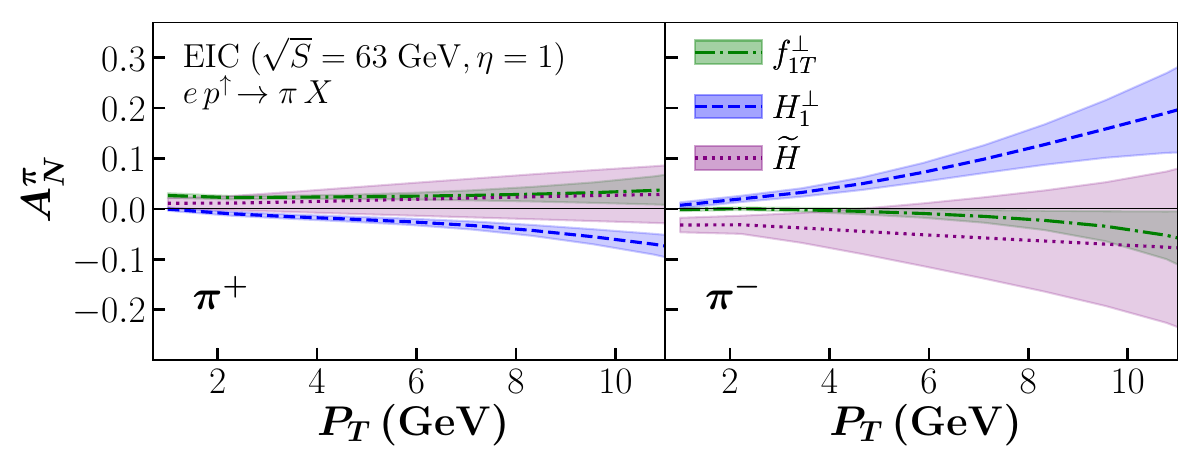}\\[0.1cm]
\includegraphics[width=0.72\textwidth]{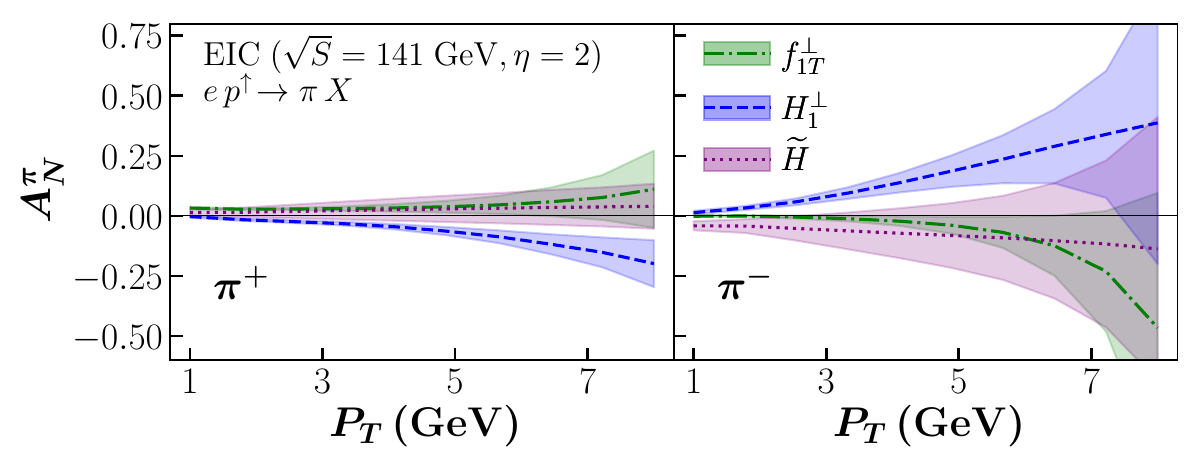} \vspace{-0.4cm}
\end{center}
\caption{Plot showing the LO central value and 1-$\sigma$ uncertainty bands of the contributions to $A_N$ in $ep^\uparrow\to \pi X$ ($\pi^+$ ($\pi^-$) left (right) panel of each subfigure) from the terms in Eq.~(\ref{e:dsigUT}) involving $f_{1T}^\perp(x)$ (green dot-dashed curve and band), $H_1^\perp(z)$ (blue dashed curve and band), and $\tilde{H}(z)$ (purple dotted curve and band) for the same EIC configurations as Fig.~\ref{f:AN_EIC}.} 
\label{f:AN_EIC_uncert}
\end{figure}

%%%%%%%%%%%%%%%%%%%%%%%%%%%%%%%%%
%%%%%%%%%%%%%%%%%%%%%%%%%%%%%%%%%
A noticeable aspect about all our computations for $A_N$ in $\ell\,N^\uparrow \to \pi\,X$ is the wide error bands.  In order to better understand the sources of this theoretical uncertainty, in Fig.~\ref{f:AN_EIC_uncert} we separately plot the central curve and 1-$\sigma$ error band of each term ($f_{1T}^\perp,H_1^\perp$, and $\tilde{H}$) that contributes to the EIC asymmetries (at LO) shown in Fig.~\ref{f:AN_EIC}.  As alluded to previously, the twist-3 FF $\tilde{H}(z)$ was only extracted for the first time in Ref.~\cite{Gamberg:2022kdb} and is not well constrained.  This correlates to the $\tilde{H}$ term having a much larger uncertainty than the $f_{1T}^\perp$ and $H_1^\perp$ terms.  At the high-energy configuration, the $f_{1T}^\perp$ and $H_1^\perp$ terms start to also have an increased uncertainty, more comparable to that of the $\tilde{H}$ term, at larger $P_T$ due to the issues discussed at the end of Sec.~\ref{s:pred}.
%%%%%%%%%%%%%%%%%%%%%%%%%%%%%%%%%%%%%%%%%
%%%%%%%%%%%%%%%%%%%%%%%%%%%%%%%%%%%%%%%%

In our analysis of $A_N$ in $\ell\,N^\uparrow \to \pi\,X$ we have found a delicate interplay between terms, sizable theoretical errors, and an intriguing impact from including NLO corrections in the unpolarized cross section.  All of these features open up possible opportunities to gain further insight from this observable into the mechanism behind TSSAs.  The important role played by the $\tilde{H}$ term and its large uncertainty will allow us to understand more about this quark-gluon-quark FF from future experiments.    Even the fact that the Sivers-like term has a significant influence on $A_N$ in  $\ell\,N^\uparrow$ collisions is in contrast to $A_N$ in $p^\uparrow p$ scattering, which is dominated by the corresponding $H_1^\perp$ contribution~\cite{Cammarota:2020qcw,Gamberg:2022kdb}.  In addition, $A_N$ in $\ell\,N^\uparrow \to \pi\,X$ appears to be a nice testing ground for the numerical significance of NLO corrections to CT3 observables, and an understanding of this issue may be required before including current and future data in a QCD global analysis.

Moreover, there is the broader question of whether $A_N$ in $\ell\,N^\uparrow \to \pi\,X$ can be explained within the postulated universal framework for TSSAs that has been used in a variety of reactions across a wide range of kinematics~\cite{Cammarota:2020qcw,Gamberg:2022kdb,Cocuzza:2023oam,Cocuzza:2023vqs} and served as the basis for our analysis.  This observable seems to be able to provide high sensitivity to the non-perturbative functions that enter the theoretical description of TSSAs -- $f_{1T}^{\perp(1)}(x)$, $h_1(x)$, $H_1^{\perp(1)}(z)$, and $\tilde{H}(z)$ -- and its measurement at the EIC is mandatory given the machine's luminosity and lever arm in $x_F$ and $P_T$~\cite{AbdulKhalek:2021gbh}. Conversely, if we are able to in the near term further constrain the aforementioned functions through additional TSSA experiments at Jefferson Lab, COMPASS, and RHIC, then we can make more precise calculations/predictions of $A_N$ in $\ell\,N^\uparrow \to \pi\,X$.  For example, measurements of the $A_{UT}^{\sin\phi_S}$ modulation in SIDIS at COMPASS can help with extractions of $\tilde{H}(z)$ and of the Sivers and Collins effects in SIDIS at Jefferson Lab in the high-$x$ regime can help pin down the Sivers TMD and transversity PDFs in the forward kinematic region of $A_N$ that has large uncertainties (see Fig.~\ref{f:AN_EIC} bottom row).

%%%%%%%%%%%%%%%%%%%%%%%%%%%%%%%%%
%%%%%%%%%%%%%%%%%%%%%%%%%%%%%%%%%
\section{Conclusions} \label{s:concl}
%%%%%%%%%%%%%%%%%%%%%%%%%%%%%%%%%
%%%%%%%%%%%%%%%%%%%%%%%%%%%%%%%%%
We have provided predictions for $A_N$ in $\ell\, N^\uparrow \to \pi\, X$ for HERMES, JLab, COMPASS, and EIC kinematics and in the first case compared to existing measurements.  This observable is the  analogue to $A_N$ in proton-proton collisions that has been studied intensely for decades.  Collectively, the use of the LIR/EOMR-simplified formula (\ref{e:dsigUT})~\cite{Kanazawa:2015ajw}, most recent simultaneous extractions of $f_{1T}^{\perp(1)}(x)$, $h_1(x)$, $H_1^{\perp(1)}(z)$, and $\tilde{H}(z)$~\cite{Gamberg:2022kdb}, and NLO formula in the $A_N$ denominator~\cite{Hinderer:2015hra} have allowed us to carry out a state-of-the-art numerical computation for this observable in the CT3 framework.  We find improved agreement with HERMES data compared to the analysis in Ref.~\cite{Gamberg:2014eia}, especially when including NLO corrections to the unpolarized cross section, which could be an empirical indication that NLO corrections to the transversely polarized cross section are (or need to be) small (see Sec.~\ref{s:NLO} for more discussion).  Overall, there  are several benefits and possible opportunities (see Sec.~\ref{s:uncert} for more details) from future measurements of $A_N$ in $\ell\, N^\uparrow \to \pi\, X$, like testing our understanding of the origin of TSSAs and further constraining quark-gluon-quark correlation functions.

\section*{Acknowledgments}  
We thank Marc Schlegel for useful discussions about the results in Ref.~\cite{Hinderer:2015hra} and status of the full NLO calculation of $A_N$, Umberto D'Alesio for inquiring about an update to the 2014 CT3 analysis, Chris Cocuzza for creating LHAPDF tables of the JAM3D-22 functions, and Rabah Abdul Khalek for providing the LHAPDF tables of DSS14 created by Valerio Bertone.  We are also grateful to Joshua Miller for his participation at earlier stages of the project.
This work was supported by the National Science Foundation under Grants No.~PHY-2011763 and No.~PHY-2308567.

\bibliographystyle{apsrev}

\end{document}